\documentclass[aps,pre,twocolumn,amsmath,amssymb,superscriptaddress,showpacs,floatfix]{revtex4-1}
\usepackage{graphicx}

\begin{document}

\title{Pinning and unbinding of ideal polymers from a wedge corner}
\author{Raz \surname{Halifa Levi}}\email{razhalifa@gmail.com}
\affiliation{Raymond and Beverly Sackler School of Physics and
Astronomy, Tel Aviv University, Tel Aviv 69978, Israel}
\author{Yacov Kantor}
\affiliation{Raymond and Beverly Sackler School of Physics and
Astronomy, Tel Aviv University, Tel Aviv 69978, Israel}
\author{Mehran Kardar}
\affiliation{Department of Physics, Massachusetts Institute of Technology,
Cambridge, Massachusetts 02139, USA}
\date{\today}

\begin{abstract}
A polymer repelled by unfavorable interactions with a uniform flat surface
may still be pinned to attractive edges and corners. This is demonstrated
by considering adsorption of a two-dimensional ideal polymer to an attractive
corner of a repulsive wedge. The well-known mapping between the statistical
mechanics of an ideal polymer and the quantum problem of a particle in a
potential is then used to analyze the singular behavior of the unbinding
transition of the polymer. The divergence of the localization length is found
to be governed by an exponent that varies continuously with the angle (when
reflex). Numerical treatment of the discrete (lattice) version of such an
adsorption problem confirms this behavior.
\end{abstract}
\maketitle

\section{Introduction}

Absorption of polymers to surfaces is a common phenomenon, manifesting a
competition between energy gain of binding and entropy loss of
fluctuations in unbound configurations. As compromise, a polymer attached
(``anchored") by one end to the surface may decrease its energy by staying
within a finite  distance $\xi$ from the surface and frequently visiting it.
The reduction in entropy of the polymer in this {\em absorbed} state is
thus compensated by a bigger gain in energy; the balance between the two
is determined by temperature $T$. Above the {\em adsorption critical
temperature} $T_a$, the polymer depins from the surface, transitioning into
a {\it delocalized} state. Such transitions have been studied in great detail
in the literature~\cite{Eisenriegler82,Binder83,debell,Livne88,Meirovitch88,Meirovitch93,EisenrieglerBook93,Vrbova98,Rychlewski11}.

Polymers in the bulk also exist in different states, with distinct universal
characteristics~\cite{degennesSC}. Configurations of polymers in good
solvents are designated as {\em self-avoiding}, with repulsive interactions
between the monomers paramount. If the latter can be ignored, the polymers
are called {\em ideal} and frequently modeled as random walks on a lattice.
These, and other polymer types, each exhibit separate singular behavior
near the adsorption transition, characterized by distinct
exponents~\cite{JansevanRensburg15a}. Transitions of {\it ideal} polymers
have been extensively  studied due to their analogy to well known models of
quantum particles in  attractive potentials~\cite{Gennes69}. For most studies
of adsorption transitions, the analogous potential includes both attractive
areas and repulsive components, to model solid surfaces  covered by an attractive
layer. (In the absence of the repulsive part, an ideal polymer is always
absorbed to an attractive layer.)

\begin{figure}
\includegraphics[width=6 truecm]{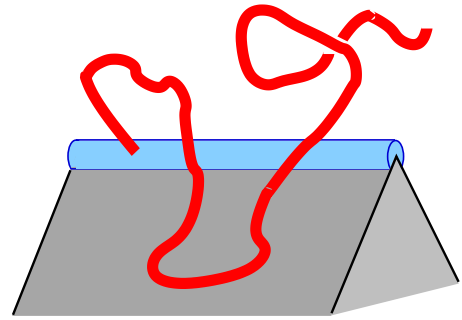}
\caption{Polymer (red) attached to a repulsive wedge (gray surface) with an
attractive edge (blue). The wedge is assumed to be infinite.}
\label{fig:PolymerOnWedge}
\end{figure}

Geometry and dimensionality play an important role in determining the
unbinding transition temperature, and its singular behavior, leading to
different characteristics for polymer adsorption to a rod~\cite{Gorbunov82,Hanke05},
a sphere~\cite{Birshtein91,Birshtein91a},  or an arbitrarily shaped
mesoscopic particles covered by an attractive layer~\cite{Hanke99}. In many
cases, the adsorbing body introduces an external {\it length scale} to the
polymer problem, through its finite size or curvature. However, there are
also interesting cases where adsorption is to a {\it scale free} form, such
as the  repulsive (infinite) wedge depicted in Fig.~\ref{fig:PolymerOnWedge},
with its edge covered by an attractive layer of microscopic diameter.
A polymer attached to such a wedge is expected to undergo an adsorption transition
with properties, including critical exponents, that depend on  dimensionless
descriptors such as the wedge angle. Such angle-dependence of critical
exponents is not new. The total partition function  of a long flexible
homogeneous polymer in a dilute solvent scales with  the number of monomers
$N$ as ${\cal Z}_{\rm tot}\sim \mu^N N^{\gamma-1}$. While the leading
(exponential) term depends on the non-universal parameter $\mu$, the exponent
$\gamma$ in the subleading power-law is universal depending on only a few major
features, such as the space  dimension $d$, or the polymer state~\cite{degennesSC}.
If a polymer is attached to a repulsive {\em scale-free} surface, such as a plane,
or the tip of a cone or a wedge, its partition function will have the same form,
but with a smaller exponent $\gamma$ due to reduction in the number of possible
configurations~\cite{Ben-Naim10,MKK_EPL96,MKK_PRE86}. Similar behavior, albeit
with a yet different set of angle-dependent exponents $\gamma$, is expected at
the desorption transition from scale-free surfaces~\cite{KK_PRE96}.

The {\it ideal} polymer in the configuration depicted in
Fig.~\ref{fig:PolymerOnWedge} maps to the quantum problem of a particle in the
two-dimensional potential obtained from a cross-section of the geometry.
As we demonstrate in this work, this problem is exactly solvable. While
different from the case of a realistic (hence self-avoiding) polymer in
three dimensions, we hope that the two share qualitative characteristics.
If the quality of a solvent is reduced, the monomers will tend to aggregate.
At the compensation point between good and poor solvent, denoted
the $\Theta$-point, the resulting polymer configurations are called
$\Theta$ polymers \cite{degennesSC}. In $d=3$ in {\em free space} many
of their characteristics are close to ideal polymers. However, we do not
expect this similarity to extend to adsorption to a line.
(The two-dimensional problem of a self-avoiding and $\Theta$ polymers
in a similar potential does not have a bound state due to the screening of
the attractive point.)
We expect {\em self-avoiding polymers} and ideal polymers in the setup of
Fig.~\ref{fig:PolymerOnWedge} to share the property of continuous variations of
exponents of the unbinding transition with the wedge angle, although
the actual exponents will naturally differ.

The remainder of the paper is organized as follows: In
Sec.~\ref{sec:polymer_quantum} we recount the analogy between a polymer in
the presence of a weak, slowly-varying, potential, and the quantum mechanical
problem of a particle in a potential well. We also formulate the problem on a
lattice and point out the differences between continuous and discrete systems.
In Sec.~\ref{sec:circular} we consider a two-dimensional problem of a
circular well confined by the repulsive walls of a wedge. We find the
critical strength of the well potential as a function of the wedge opening
angle, and characterize the singular behavior of the unbinding transition.
The dependence on polymer length $N$, and the exponent $\gamma$ are
detailed in Sec.~\ref{sec:Ndependent}. The discrete version of the problem,
with an attractive lattice site located near a wedge is described in
Sec.~\ref{sec:point}, where we determine numerically both the transition point
and the correlation length exponent for several wedge angles
(Sec.~\ref{sec:LatticeCritical}).

\section{Analogy to quantum bound states}\label{sec:polymer_quantum}

The well-known mapping between adsorption of a polymer and bound states in
quantum mechanics~\cite{Gennes69} is briefly reviewed here. Let
${\cal Z}({\bf r}, {\bf r}_0,N)$ denote the partition function
of an ideal polymer of $N$-steps, of mean squared  size $\ell^2$,
that starts at point ${\bf r}_0$ and ends at point ${\bf r}$. In
free space the total partition function is
${\cal Z}_0\equiv\int {\cal Z}{\rm d}^d{\bf r}=\mu^N$, and it
is convenient to define the {\em reduced} partition function
$\tilde{\cal Z}({\bf r}, {\bf r}_0,N)={\cal Z}({\bf r}, {\bf r}_0,N)/{\cal Z}_0$.
If the potential affecting the monomers, $V^{\rm th}({\bf r})$, changes slowly,
such that at temperature $T$ its change over the distance $\ell$ is much
smaller than $k_BT\equiv \beta^{-1}$, then the partition function difference
$\tilde{\cal Z}({\bf r}, {\bf r}_0,N+1)-\tilde{\cal Z}({\bf r}, {\bf r}_0,N)
\approx \partial \tilde{\cal Z}/\partial N$ can be cast in the continuum
form~\cite{Wiegel86}
\begin{equation}\label{eq:Ndepend}
\frac{\partial \tilde{\cal Z}}{\partial N}=
\frac{\ell^2}{2d}\nabla^2\tilde{\cal Z}-\beta V^{\rm th}\tilde{\cal Z},
\end{equation}
supplemented  with the initial condition $\tilde{\cal Z}({\bf r}, {\bf r}_0,0)=\delta^d({\bf r}-{\bf r}_0)$.
This equation can be solved by variable separation, which leads to
eigenvalue equation
\begin{equation}\label{eq:th-eigenvalue}
\left(-\frac{\ell^2}{2d}\nabla^2+\beta V^{\rm th}\right)f_\alpha=E^{\rm th}_\alpha f_\alpha.
\end{equation}
Knowledge of all the eigenfunctions $f_\alpha$, and their eigenvalues (``energies")
$E^{\rm th}_\alpha$, enables reconstruction of the reduced partition function as
\begin{equation}\label{eq:Zdecomposition}
\tilde{\cal Z}({\bf r}, {\bf r}_0,N)=\sum_\alpha f_\alpha({\bf r})f^*_\alpha({\bf r}_0)
{\rm e}^{-E^{\rm th}_\alpha N}.
\end{equation}
The analogy of the above treatment with the single particle Schr{\"o}dinger
equation is immediately apparent. In this analogy,  the variable $N$
corresponds to an imaginary time for the quantum particle, its  mass $m$ and
potential $V^{\rm q}$ related by $\beta V^{\rm th}d/\ell^2=m V^{\rm q}/\hbar^2$,
with the same scaling for $E_\alpha$s in the eigenvalue equation.

If the potential $V^{\rm th}({\bf r})$ includes  attractive
parts, it may support bound states~\cite{Buell95} with discrete eigenvalues
$E^{\rm th}_\alpha<0$. If there is a gap between the ground and the first
 excited state, for large $N$ the solution will be dominated by the
ground state ($\alpha=0$), and
\begin{equation}
\tilde{\cal Z}({\bf r}, {\bf r}_0,N)\approx f_0({\bf r})f_0({\bf r}_0)
{\rm e}^{-E^{\rm th}_0 N}.
\end{equation}
Since $\tilde{\cal Z}$ is positive, the ground state function $f_0({\bf r})$
cannot alternate in sign, and can be chosen as being non-negative everywhere.
(The absence of nodes in the ground state of a quantum particle is well known.)
A bound state $f_0({\bf r})$ will be localized within some localization length $\xi$ in
the neighborhood of the well. Since $\tilde{\cal Z}$ is proportional to the
probability to find the polymer end at $\bf r$, this implies that the polymer
is also localized in the vicinity of the attractive potential. Assuming a
typical linear size $a$ of the potential ``well," it is convenient to recast
the equation in terms of dimensionless coordinates ${\bf r}'={\bf r}/a$, as
\begin{equation}\label{eq:DimensionlessEigenvalue}
\left(-\nabla'^2+V\right)f_\alpha=E_\alpha f_\alpha.
\end{equation}
Here, $\nabla'^2$ represents the Laplacian in dimensionless coordinates,
$E_\alpha\equiv(2a^2d/\ell^2)E^{\rm th}_\alpha$ are the dimensionless energy
eigenvalues, and
\begin{equation}\label{eq:correspondence}
V\equiv\frac{2d\beta a^2}{\ell^2}V^{\rm th},
\end{equation}
is the dimensionless potential. For further reference, we note that
the $N$-dependent Eq.~\eqref{eq:Ndepend} in the new dimensionless
variables, can be expressed as
\begin{equation}\label{eq:NdependDimensionless}
\frac{\partial \tilde{\cal Z}}{\partial N'}=
\nabla'^2\tilde{\cal Z}-V\tilde{\cal Z},
\end{equation}
where $N'\equiv N{\ell^2}/(2da^2)$. Note that $E_\alpha N'=E^{\rm th}_\alpha N$.
In what follows, we omit the prime in coordinate notation and always measure
the distances relative to the extent of the potential.

It is well known in quantum mechanics that any purely attractive potential
in $d=1$ dimension always has a bound state~\cite{LL_vol3}, while a
sufficiently deep well may have many bound states. (There is also a slightly
more relaxed criterion guaranteeing the presence of bound
states~\cite{Brownstein00}.) The situation is similar in $d=2$ dimensions,
where a bound state can always be found~\cite{Chadan03}. As a concrete
example, consider a circular well of unit radius
\begin{equation}\label{eq:Vcirc}
V_{\rm circ}({\bf r})=
\begin{cases}
-V_0, & \text{for } r<1 \\
\ \ \ 0 ,&   \text{for } r\ge 1
\end{cases}.
\end{equation}
The above discontinuous potential was chosen for its simplicity, since we expect that
the universal features of the unbinding transition are independent of its detailed shape
(as is the case in $d=1$).
This choice may appear to contradict the statement at the beginning of this Section
that the analogy of the ideal polymer to the  ``quantum particle" in Eq.~\eqref{eq:Ndepend}
is valid only for slowly varying potentials.
The actual requirement is that  for a potential $V^{\rm th}\sim k_BT$ the range $\Delta r$ of the change in the
potential should satisfy $\ell\ll\Delta r$. The discontinuous potential in
Eq.~\eqref{eq:Vcirc} can thus be viewed as the continuum limit of the case of $\ell\ll\Delta r\ll a=1$,
and therefore represents a valid situation for this mapping.

The eigenfunctions in both interior and exterior of the well described
by Eq.~\eqref{eq:Vcirc}  are Bessel
functions. In case of the ground state they correspond to the regular
and second modified Bessel functions $J_0$ and $K_0$, respectively.
For a shallow well ($V_0\ll 1$) the ground state energy is extremely
small ($E_0\sim {\rm e}^{-4/V_0}$), and the corresponding localization
length is very large~\cite{Nieto02}. In higher dimensions $d$, the
presence or absence of bound states depends on the depth and details
of the potential. In fact, if $d$ is viewed as continuous
variable, it can be shown~\cite{Nieto02} that the property of
always having a bound state disappears immediately above $d=2$.

The above theorems do not apply to potentials that have both repulsive and
attractive parts. For instance, a one-dimensional potential representing an
attractive layer on a repulsive wall may have no bound states if it is
shallow enough. We shall see that a similar situation appears for
a two-dimensional circular well in the presence of repulsive walls.

Many theoretical studies of  polymers near attractive and repulsive
surfaces are performed on discrete lattice models. We will consider a
$d$-dimensional hypercubic lattice, with lattice spacing $\ell$, with
polymer configurations represented by  $N$-step walks. The total partition
function of a polymer in the absence of any potentials is ${\cal Z}_0=(2d)^N$.
The potential $V^{\rm th}$ is modeled by Boltzmann weights
$q({\bf r})=\exp(-\beta V^{\rm th})$ assigned to lattice sites. In free
space $q=1$,  on the repulsive wall $q=0$, while for well of depth
$V^{\rm th}=-V^{\rm th}_0$, $v=\exp(\beta V^{\rm th}_0)$. The reduced
$(N+1)$-step partition function can be deduced from $N$-step reduced
partition function by
\begin{equation}\label{eq:ZNdiscrete}
\tilde{\cal Z}({\bf r},{\bf r}_0,N+1)=\frac{q({\bf r})}{2d}
\sum_{{\bf r}'\text{ nn of }{\bf r}}
\tilde{\cal Z}({\bf r}',{\bf r}_0,N),
\end{equation}
with the starting condition
$\tilde{\cal Z}({\bf r},{\bf r}_0,0)=q({\bf r}_0)\delta_{{\bf r},{\bf r}_0}$.

As an aside, note that hypercubic lattices are bipartite, their sites separable
into disjoint  ``even" (``e") and ``odd" (``o") subsets. A walk starting on one
subset lands on the same subset (or its complement) after an even (odd) number
of steps, with $\tilde{\cal Z}({\bf r},{\bf r}_0,N)=0$ on one or other subset.
Equation \eqref{eq:ZNdiscrete} can be written in matrix form
\begin{equation}\label{eq:matrix}
\tilde{\cal Z}_{N+1}=M \tilde{\cal Z}_{N},
\end{equation}
where the matrix $M({\bf r},{\bf r}')$ connects ``e" and ``o" coordinates.
A further recursion yields
\begin{equation}\label{eq:matrix2}
\tilde{\cal Z}_{N+2}=M^2 \tilde{\cal Z}_{N},
\end{equation}
which connects only sites of the same type. Thus the matrix $M^2$ can be
decomposed into two completely unconnected sub-matrices. Since the
elements of each sub-matrix are positive, the largest eigenvalue
$\lambda^2$ is real positive, and the corresponding  eigenvector is unique.
Indicating by $\psi_{\rm e}({\bf r})$ the eigenvector in the even subspace,
\begin{equation}\label{eq:DiscreteEigen}
\lambda^2\psi_{\rm e}({\bf r})=M^2({\bf r},{\bf r}')\psi_{\rm e}({\bf r}'),
\end{equation}
it is easy to see that
$\psi_{\rm o}({\bf r})\equiv M({\bf r},{\bf r}')\psi_{\rm e}({\bf r}')$
is an eigenvector (with eigenvalue $\lambda^2$) in the odd subspace.
The orthogonal states $\psi_{\rm o}$ and $\psi_{\rm e}$ are in the above sense
``ground states" of the problem, with energy $E_0$ obtained from $\lambda^2={\rm e}^{-2E_0}$.

For a weak potential with small variations between adjacent lattice
sites, Eq.~\eqref{eq:ZNdiscrete} coincides with Eq.~\eqref{eq:Ndepend}. However,
in a typical lattice simulation the geometrical features are reduced to
a bare minimum, e.g., an attractive well near a flat repulsive surface
is represented by a {\em single} lattice layer of sites with some weight $w$,
while a well near a repulsive wall or wedge may be represented by a single
attractive site with a weight $v$. Since the width of the attractive
layer, or a well $a$, coincide with the polymer step size $\ell$, we may
expect only qualitative similarity between the solutions of
Eqs.~\eqref{eq:Ndepend} and \eqref{eq:ZNdiscrete}. Nevertheless, we shall
see that the results of continuous and discrete problems are rather
close in their numerical values.

\section{Attractive circular well within a sector}\label{sec:circular}

\begin{figure}
  \includegraphics[width=6 truecm]{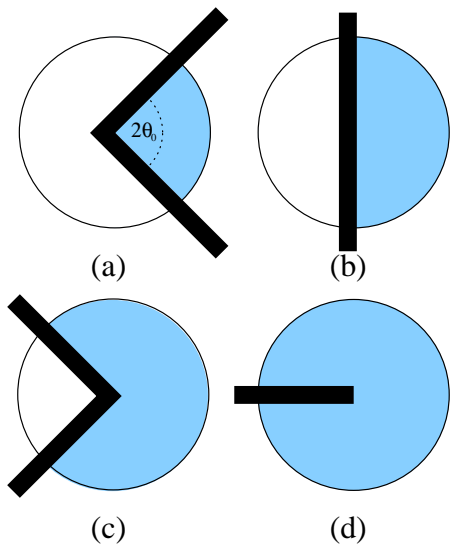}
\caption{Circular potential well (blue area) of radius 1
and depth $-V_0$, confined to a sector of opening angle
$2\theta_0$ by hard walls. The sub-figures represent examples of
(a) inside a rectangular sector ($\theta_0=\pi/4$),
(b) near a flat surface (line) ($\theta_0=\pi/2$),
(c) outside a rectangular sector ($\theta_0=3\pi/4$), and
(d) outside a needle-like sector ($\theta_0=\pi$).
}
\label{fig:sector}
\end{figure}

We next consider a circular well, as in Eq.~\eqref{eq:Vcirc}, near a repulsive
wall, either flat or forming a sector with opening angle $2\theta_0$ as depicted
in Fig.~\ref{fig:sector}. The potential will be
\begin{equation}\label{eq:sector}
V_{\rm{sect},\theta_0}(r,\phi)=\begin{cases}
V_{\rm circ}({\bf r}), & \text{for } -\theta_{0}<\phi<\theta_{0},\\
\infty, & \text{otherwise,}
\end{cases}
\end{equation}
where $r$ and $\phi$ are the polar coordinates depicting the distance
from the origin, and the azimuthal angle, respectively. As before, we
assume that the distances are measured relative to the circle radius.

In polar coordinates, Eq.~\eqref{eq:DimensionlessEigenvalue} for potential
$V_{\rm sect}$ becomes
\begin{equation}\label{eq:polar}
-\left[\frac{1}{r}\frac{\partial}{\partial r}\left(r\frac{\partial f_\alpha}{\partial r}\right)+\frac{1}{r^{2}}\frac{\partial^{2}f_\alpha}{\partial\phi^{2}}\right]=\epsilon_\alpha f_\alpha,
\end{equation}
where $\epsilon_\alpha \equiv E_\alpha+V_0 $ inside the well ($r<1$),  and
$\epsilon_\alpha \equiv E_\alpha$ outside the well ($r>1$). The repulsive
walls of the sector are enforced by the boundary condition
$f_\alpha(r,\phi=\pm\theta_0)=0$. The eigenfunction $f_\alpha$ and its
derivative are continuous at $r=1$. To ensure that the function vanishes on
the boundaries of the sector, its angular dependence in constrained to the forms
$\sin(\nu\phi)$, with $\nu=(\pi/\theta_0)n$, for $n=1,2,\cdots$; or
$\cos(\nu\phi)$, with $\nu=(\pi/\theta_0)(n+\frac{1}{2})$, for $n=0,1,2,\cdots$.
In general, $\nu$ is not an integer, except at $\theta_0=\pi/2$, when the sector
becomes a flat surface. Discrete eigenstates of a circular well with
{\em infinite} walls, i.e., with $V=\infty$ for $r>1$, confined by a sector were
analyzed in detail in Ref.~\cite{Robinett03}. Our problem of a well of {\em finite}
depth admits both discrete ($E_\alpha<0$) and continuous ($E_\alpha\ge 0$) spectra.
For $E_\alpha<0$ the radial part of the solution is~\cite{NISTlib} a regular Bessel
function $J_\nu(kr)$ with $k=\sqrt{E_\alpha+V_0}$ for $r<1$, and second modified
Bessel function $K_\nu(qr)$ with $q=\sqrt{-E_\alpha}$ for $r>1$. This  choice of
the  Bessel functions ensures regularity of the solution at $r=0$, and its
vanishing  for $r\to\infty$. Since Eq.~\eqref{eq:DimensionlessEigenvalue} is a
{\em second order} linear equation, for {\em finite} potential $V$, the second
derivative must exist and consequently both the function and its derivative must
be continuous. In the presence of the finite jump in $V$ there is a {\em finite}
jump in the {\em second} derivative, but the first derivative remains continuous
as in the case of smooth $V$. Therefore, the eigenvalue $E_\alpha$, should be
selected to enforce continuity of the function and its derivative at $r=1$. Thus
$E_\alpha$ is a solution of the equation
\begin{equation}\label{eq:continuity}
\frac{kJ'_\nu(k)}{J_\nu(k)}=\frac{qK'_\nu(q)}{K_\nu(q)},
\end{equation}
where the prime denotes derivative of a function with respect to its argument.
For large $V_0$ we can have multiple bound states for several values of $\nu$.

\begin{figure}[t!]
\includegraphics[width=8.5 truecm]{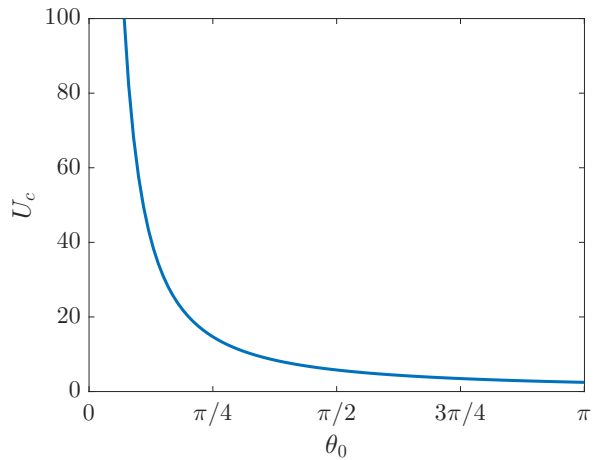}				
\caption{Critical depth of a well $U_c$ as a function
of the half of the central angle $\theta_0$ of the sector.}
		\label{fig:lnUvsQ}
\end{figure}

Since we are interested in the ground state, we will consider only
$\nu=m\equiv\pi/2\theta_0$, since larger $\nu$ solutions have alternating
signs. Of all possible solutions for this $m$ we look for the smallest
$k$. The number of bound states will decrease with decreasing $V_0$, and
for a limiting $V_0=U_c$ the eigenenergy $E_0$ of the ground state is zero,
reducing Eq.~\eqref{eq:continuity} to
\begin{equation}\label{eq:circUc}
\frac{\sqrt{U_c}J'_m(\sqrt{U_c})}{J_m(\sqrt{U_c})}
=\lim_{q\to0}\frac{qK'_m(q)}{K_m(q)}=-m.
\end{equation}
By using the recurrence relations of Bessel functions and their
derivatives~\cite{Abramowitz72}, we find that Eq.~\eqref{eq:circUc} is
reduced to solving $J_{m-1}\left(\sqrt{U_{c}}\right)=0$. Thus, $U_c$ is
simply the square of the first zero of $J_{(\pi/2\theta_0)-1}$.
Figure~\ref{fig:lnUvsQ} depicts the dependence of $U_c$ on the angle
$\theta_0$. As expected, the critical depth $U_c$ diverges with
increasing confinement for $\theta_0\to0$. Since for large $m$ the first
root of $J_m$ is approximately at $m$~\cite{Laforgia84}, for small
$\theta_0$ we have $U_c\approx m^2=(\pi/2\theta_0)^2$. At the other
extreme $U_c(\theta_0=\pi)=\pi^2/4$, while the radial part of the ground
state eigenfunction  at $E_0=0$ inside the well becomes
$J_{1/2}(\pi r/2)$, where $J_{1/2}(x)\sim\sin(x)/\sqrt{x}$. Thus $U_c$
remains finite when the sector becomes a needle-like insertion into the
attractive well and the eigenfunction has a very different shape from
the bound states of a circular well without the repulsive walls. For
$\theta_0=\pi/2$ the sector becomes a straight line, and $U_c\approx 5.78$.

For $V_0>U_c$ bound states are present, and the
partition function at large distances is described by
$K_{m}(qr)\sim\sqrt{\frac{\pi}{2qr}}{\rm e}^{-qr}$. The exponential decay of
this function implies that the polymer is localized in the vicinity
of the corner over a distance  of order $\xi=1/q=1/\sqrt{-E_0}$.
(For long polymers, only the ground state needs to be taken into account.)
When $E_0=0$ the exponential long distance decay of $K_{m}(qr)$ is replaced
by a power law $r^{-m}$. This is different from the one-dimensional case, where
$E_0=0$  corresponds to a function that is constant outside the well. When the
depth of the well is close to $U_c$, i.e. for small $\delta V_0\equiv V_0-U_c$,
$E_0$ is also  small, and we can expand the left hand side of
Eq.~\eqref{eq:continuity} (with $\nu=m$)
${\cal F}_{{\rm L},m}(k)={\cal F}_{{\rm L},m}(\sqrt{U_c+\delta V_0+E_0})\equiv
kJ'_m(k)/J_m(k)$ in $\delta V_0+E_0$
\begin{equation}
{\cal F}_{{\rm L},m}\approx -m-\frac{1}{2}(E_0+\delta V_0),
\end{equation}
and also expand the right hand side of the same equation
${\cal F}_{{\rm R},m}(q)={\cal F}_{{\rm R},m}(\sqrt{-E_0})\equiv
qK'_m(q)/K_m(q)$ in $(-E_0)$. The latter expansion  has different
forms depending on the value of $m=2\theta_0/\pi$~\cite{Abramowitz72}:

\begin{equation}
{\cal F}_{{\rm R},m}\!\approx\!
\begin{cases}
-m-\frac{1}{2(m-1)}(-E_0),  &\text{for } m>1,\\
-1+\frac{1}{2}(-E_0)\ln(-E_0), &\text{for } m=1,\\
-m-\frac{2^{1-2m}\pi}{\sin\left(\pi m\right)\Gamma^{2}\left(m\right)}\left(-E_0\right)^m\!\!, &
\text{for }\frac{1}{2}\leq m<1.
\end{cases}
\end{equation}
By  equating ${\cal F}_{{\rm L},m}={\cal F}_{{\rm R},m}$ we
find that, to the leading order,  $E_0$ depends on $\delta V_0$ as
follows
\begin{equation}
q^2=-E_0\sim
\begin{cases}
\delta V_0, &\text{for } 0<\theta_0<\frac{\pi}{2},\\
\delta V_0/|\ln \delta V_0|, &\text{for } \theta_0=\frac{\pi}{2},\\
\delta V_0^{2\theta_0/\pi}, &\text{for } \frac{\pi}{2}<\theta_0\leq\pi.
\end{cases}
\end{equation}
Since the dimensionless potential $V_0$ depends on the physical potential
$V_0^{\rm th}$ and the temperature via Eq.~\eqref{eq:correspondence},
$\delta V_0=(2a^2d/k_B\ell^2)V_0^{\rm th}(1/T-1/T_a)\approx
(2a^2d/k_BT_a^2\ell^2)V_0^{\rm th}(T_a-T)$, we can relate the divergence
of the localization length to the temperature difference by
\begin{equation}
\xi\sim\begin{cases}
(T_a-T)^{-\frac{1}{2}}, &\text{for } 0<\theta_0<\frac{\pi}{2},\\
\left\vert\ln(T_a-T)\right\vert^{\frac{1}{2}}(T_a-T)^{-\frac{1}{2}},
&\text{for } \theta_0=\frac{\pi}{2},\\
(T_a-T)^{-\frac{\theta_0}{\pi}}, &\text{for } \frac{\pi}{2}<\theta_0\leq\pi.
\end{cases}\label{eq:xiVsT}
\end{equation}
In the above equation, $T_a$  itself depends on $\theta_0$ due to the
$\theta_0$-dependence of the critical depth $U_c$. The $\theta_0\to\pi$
limit in the third case of Eq.~\eqref{eq:xiVsT} represents a needle-like
insertion into the circular well and differs from the situation when the
walls of the wedge are completely absent, since the semi-infinite straight
line presents a significant obstacle for a random walk. The dependence of
the critical behavior in Eq.~\eqref{eq:xiVsT} is reminiscent of other
power-law dependencies, such as that of the density of the polymer, or the
pressure it exerts on walls of a wedge~\cite{HK_PRE89}. (In particular, see
Fig.~5 in Ref.~\cite{HK_PRE89}.) They are all manifestations of scale-invariance
of the geometry, and appropriate polymer properties. Detailed knowledge of
the pressure distribution may shed light on the non-trivial behavior
described by  Eq.~\eqref {eq:xiVsT}.  The results in Ref.~\cite{HK_PRE89}
are expected to be valid in the desorbed phase.  The formalism presented in this
paper does not allow direct calculation of the local pressure distribution.
It is likely that other existing analytical and numerical, continuous space and lattice methods~\cite{Bickel01,Breidenich2000,Jensen13,HK_JCP141,HK_PRE89} can be extended to
calculations of pressure in the presence of attractive potentials.

Close to the transition point, and at very large distances $r\gg\xi$, the bound
eigenstate has radial component $K_m(r/\xi)\sim {\rm e}^{-r/\xi}\sqrt{\xi/r}$.
However, for distances $a<r\ll\xi$, where $a=1$ is the well radius  in our
calculations, the eigenstate has a simple power law dependence $\sim r^{-m}$.
The `typical' polymer size  is usually determined through an average of a
power of end-to-end distance, such as  $R^n\equiv\langle r^n\rangle$.
However, whether or not this quantity reflects the localization length $\xi$
 depends on the powers $n$ and $m$. The value of $m$ also determines
whether the relevant normalization is determined by $\xi$ or $a$.
Thus, for various cases we find
\begin{equation}
R^n\sim\begin{cases}
\xi^n, &\text{for } m<2,\\
\xi^n/\ln(\xi/a), &\text{for } m=2,\\
\xi^{2-m+n}/a^{2-m}, &\text{for } 2<m<2+n,\\
a^n\ln(\xi/a), &\text{for } m=2+n,\\
a^n, &\text{for } m>2+n.
\end{cases}\label{eq:Rn}
\end{equation}
Close to the transition, the correlation length will be
very large, and a sufficiently long polymer settles into the localized
ground state, with measures of its size given by Eq.~\eqref{eq:Rn}. However,
for moderate values of $N$, the partition function will evolve as a Gaussian,
i.e.,it  will have width of order of $\sqrt{N}$. In such a case, $\xi$ in the
above expressions should be replaced by $\sqrt{N}$.

\section{$N$-dependent solutions inside the sector}\label{sec:Ndependent}

At very high temperatures the dimensionless potential $V$ in
Eq.~\eqref{eq:NdependDimensionless} becomes negligible, reducing it to the
simple diffusion equation
\begin{equation}\label{eq:Diffusion}
\frac{\partial \tilde{\cal Z}}{\partial N'}=
\nabla^2\tilde{\cal Z},
\end{equation}
where the rescaled polymer length $N'$ plays the role of time. The required
solution for $\tilde{\cal Z}$ must still vanish on the {\em repulsive}
boundaries of the sector, corresponding to {\em absorbing} boundaries for the
diffusers. The exact solution to this problem depends on the initial position
of the diffuser (starting point of the polymer) ${\bf r}_0$. However, for
sufficiently large  $N'\gg r_0^2$, a significant portion of diffusing density
will reach the boundary, and the memory of the initial condition is only
reflected in the prefactor of the asymptotic solution~\cite{HK_PRE89,AK_PRE91}
\begin{equation}\label{eq:NdepSolution}
\tilde{\cal Z}({\bf r_0},{\bf r},N')=c\frac{r^m}{N'^{1+m}}
        {\rm e}^{-r^2/4N'}\cos(m\phi).
\end{equation}
As before, the angle $\phi$ is  measured form the axis of symmetry and
$m=\pi/2\theta_0$, while the prefactor $c$ depends on ${\bf r_0}$.
Integration of the reduced partition function leads to total partition function
$\tilde{\cal Z}_{\rm tot}={\rm const}\cdot N'^{-m/2}$, and therefore exponent
$\gamma=1-m/2=1-\pi/4\theta_0$~\cite{Ben-Naim10,MKK_EPL96,MKK_PRE86}.

The search for an $N'$-dependent solution of the diffusion equation
in a wedge with absorbing boundaries (corresponding to repulsive surfaces
for the polymer) revealed~\cite{HK_PRE89,AK_PRE91}, that besides the
solution in Eq.~\eqref{eq:NdepSolution} there is a complementary solution
\begin{equation}\label{eq:NdepSolutionComp}
\tilde{\cal Z}({\bf r_0},{\bf r},N')=\frac{c}{r^mN'^{1-m}}
        {\rm e}^{-r^2/4N'}\cos(m\phi),
\end{equation}
where the notations are the same as before. This, solution is not appropriate
for a purely absorbing wedge, since it diverges as $r^{-m}$ for $r\to0$,
and was thus discarded for the  repulsive boundary problem in
Refs.~\cite{HK_PRE89,AK_PRE91}.  However, we note that such functional form
resembles the ground state at the transition point for a polymer in a wedge
with an attractive well. Unlike the true ground state, this solution has an
$N'$-dependent prefactor and a Gaussian function that truncates the power law
behavior, and is a candidate for the asymptotic $N'$-dependent solution for the
reduced partition function outside the well (valid for sufficiently large $N'$,
when the details of the initial condition have been forgotten). We do not know
the continuation of this function inside the well, but expect that it can be
constructed by superposition of states. The fact that in the small $r$ limit the logarithmic derivative
$d\ln\tilde{\cal Z}/d\ln r\to-m$ indicates that this might be a general solution
of a small and deep attractive region in the wedge at the transition temperature.
We shall later verify this assumption by numerical solutions of discrete problems.

\begin{figure}[t!]
\includegraphics[width=8 truecm]{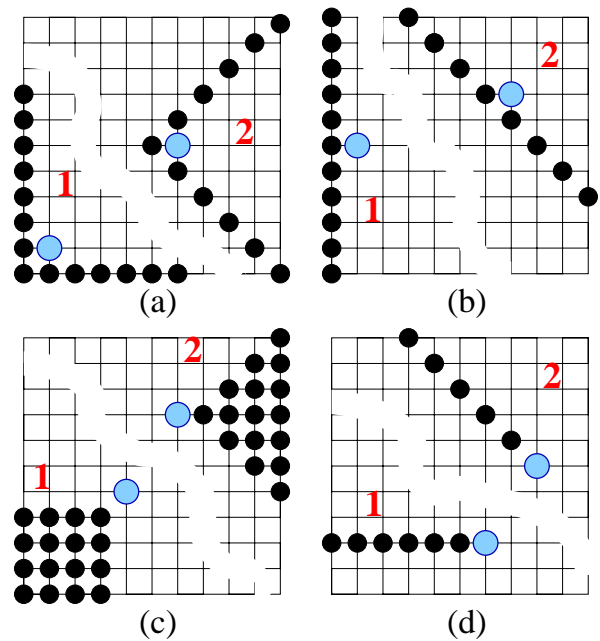}
\caption{Illustration of a single adsorbing point (light blue circle
with Boltzmann weight $v>1$), at corner of a repulsive  sector (black full circles
with weight 0) on a square lattice. For each sector
semi-angle $\theta_0$, two discrete variants of the continuous geometry
(corresponding pictures in Fig.~\ref{fig:sector})
are shown: ``1" corresponds to straight lines along the lattice
axes, and ``2" - along the lattice diagonals. The number of
repulsive sites neighboring the adsorbing site, $n$, varies between the
two variants.: In panel (a) the attracting point inside the rectangular repulsive sector ($\theta_0=\pi/4$)
has $n_1=2$ nearest neighbor repulsive sites in variant ``1", and $n_2=3$ in ``2".
In panel (b), for the attracting point near a flat repulsive surface (line) ($\theta_0=\pi/2$),
 $n_1=1$ and $n_2=2$.
In panel (c), for the attracting point outside a rectangular repulsive sector ($\theta_0=3\pi/4$),
 $n_1=0$ and $n_2=1$.
In panel (d), an attracting point at the tip of a semi-infinite repulsive line
($\theta_0=\pi$), has  $n_1=1$ and $n_2=0$.}
\label{fig:DiscreteGeometries}
\end{figure}

Assuming that Eq.~\eqref{eq:NdepSolutionComp} indeed correctly represents
the reduced partition function, we can integrate the expression to
obtain the total partition function
$\tilde{\cal Z}_{\rm tot}={\rm const}\cdot N'^{\gamma-1}$, with
\begin{equation}
\gamma=\begin{cases}
1+m/2, &\text{for } m<2,\\
2 \text{ (with log correction),} &\text{for } m=2,\\
m, &\text{for } m>2.
\end{cases}\label{eq:gamma}
\end{equation}
These exponents are larger than 1, the value for free space, and therefore
the mixture of repulsive wedge with critical attractive point at its corner
corresponds to an overall attraction for a polymer in free space.

\section{Attractive point near a repulsive sector on a
lattice}\label{sec:point}

For numerical studies of polymer adsorption,  discrete (lattice) models
provide convenient realizations. Some simple discrete analogs for a small
adsorbing well at the corner of a repulsive wedge are depicted in
Fig.~\ref{fig:DiscreteGeometries}, with the attractive potential acting on
a single point near a  line of repulsive sites. Equation~\eqref{eq:ZNdiscrete}
provides a simple recursive numerical tool for calculating
$\tilde{\cal Z}({\bf r},{\bf r}_0,N+1)$ in terms of
$\tilde{\cal Z}({\bf r},{\bf r}_0,N)$: Starting with
$\tilde{\cal Z}({\bf r},{\bf r}_0,0)=q({\bf r}_0)\delta_{{\bf r},{\bf r}_0}$,  Eq.~\eqref{eq:ZNdiscrete}
is iterated $N$ times for polymer length $N$. The resulting
$\tilde{\cal Z}({\bf r},{\bf r}_0,N)$ is proportional to the probability of
finding the end of the polymer at  ${\bf r}$. The total reduced partition
function
$\tilde{\cal Z}_{\rm tot}({\bf r}_0,N)=\sum_{\bf r}\tilde{\cal Z}({\bf r},{\bf r}_0,N)$
is the normalizing factor for this probability.

\begin{figure}[t!]
\includegraphics[width=8 truecm]{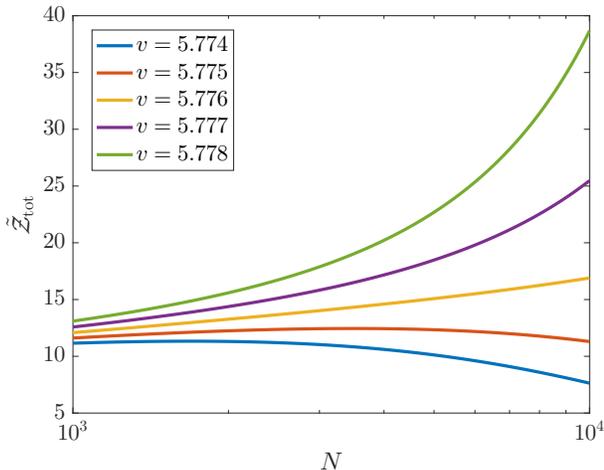}
\caption{Semi-logarithmic plot of the dependence of the total reduced partition function
$\tilde{{\cal Z}}_{\rm tot}$ on polymer length $N$
for an attractive point located inside a rectangular wedge ($\theta_0 = \pi/4$), as
depicted in Fig.~\ref{fig:DiscreteGeometries}(a) for geometry ``1", for
Boltzmann weights $v$ ranging from 5.774 to 5.778 (bottom to top).}
\label{fig:Z1}
\end{figure}

The total reduced partition function $\tilde{\cal Z}_{\rm tot}({\bf r}_0,N)$
is, by definition equal to one in free space. In the absence of attractive
potential, a polymer near a repulsive sector will have partition function
reduced to $\tilde{\cal Z}_{\rm tot}\sim N^{\gamma-1}$, where
$\gamma=1-\pi/4\theta_0<1$ depends on the  wedge
angle~\cite{Ben-Naim10,MKK_EPL96,MKK_PRE86}. This behavior persists for a
weakly attractive potential, with $v$  close to 1. As $v$ increases towards
$v_c$, this scaling is delayed to larger values of $N$.
In the adsorbed state $\tilde{\cal Z}_{\rm tot}$ will increase exponentially
with $N$ due to extra Boltzmann weights gained upon  repeated returns to the
attracting point. Again, the exponential growth, immediately apparent for $v\gg v_c$,
is delayed to larger $N$ as $v$ is decreased towards $v_c$. Figure~\ref{fig:Z1}
depicts the dependence of $\tilde{\cal Z}_{\rm tot}$ for several values of
$v$ very close to $v_c$. Only for polymers of several thousand steps does it become
evident that the two lowest graphs correspond to delocalized states,
while the two top graphs, and possibly the middle one, represent
adsorbed states.

\begin{figure}[t!]
\includegraphics[width=8 truecm]{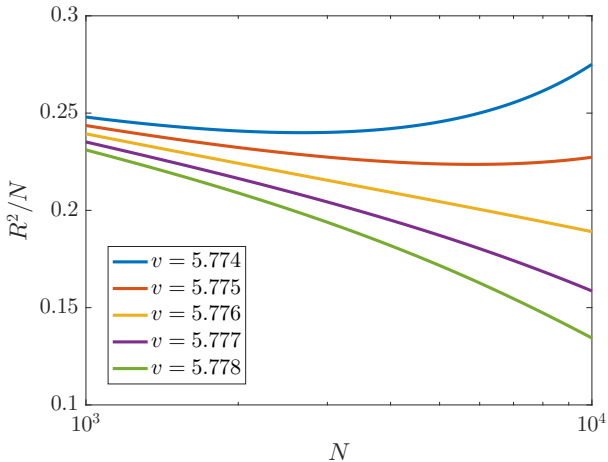}
\caption{Semi-logarithmic plot of the scaled squared end-to-end distance
$R^2$, as a function of polymer length $N$ for an attractive
point located inside a rectangular wedge ($\theta_0 = \pi/4$), as depicted
in Fig.~\ref{fig:DiscreteGeometries}a for geometry ``1", with
Boltzmann weights $v$ ranging from 5.774 to 5.778 (top to bottom).}
\label{fig:R1}
\end{figure}

If ${\bf r}_0$ is located at or near the attractive point, we can use the
mean squared end-to-end distance
\begin{equation}
R^2(N)=\frac{1}{\tilde{\cal Z}_{\rm tot}({\bf r}_0,N)}
\sum_{\bf r}({\bf r}-{\bf r}_0)^2\tilde{\cal Z}({\bf r},{\bf r}_0,N),
\end{equation}
as an indicator of the localization length. In the absence of the attractive
potential, the repulsive walls  push away the polymer while maintaining
the scaling of the random walk, such that for a polymer anchored close to the
apex of a wedge $R^2=N(1+\pi/4\theta_0)$~\cite{HK_PRE89,AK_PRE91}. (In this
section  distances are measured in units of lattice spacing, corresponding to
$\ell=1$.) This is indeed the behavior observed for $v$ close to 1. In an
adsorbed state $R^2$ approaches a constant as $N$ increases, masked by crossovers
close to $v_c$. Figure~\ref{fig:R1} depicts $R^2/N$ as a function of $N$ for the
same  values of $v$ is in Fig.~\ref{fig:Z1}. The two lowest curves in this
figure, and, possibly the middle one, correspond to adsorbed state, with the
top graphs in delocalized states. From the last two figures we estimate that
$v_c\approx 5.776$. Larger $N$ will enable even more accurate determination
of $v_c$.

\begin{figure}[t!]
\hspace*{3mm}
\includegraphics[width=8.7 truecm]{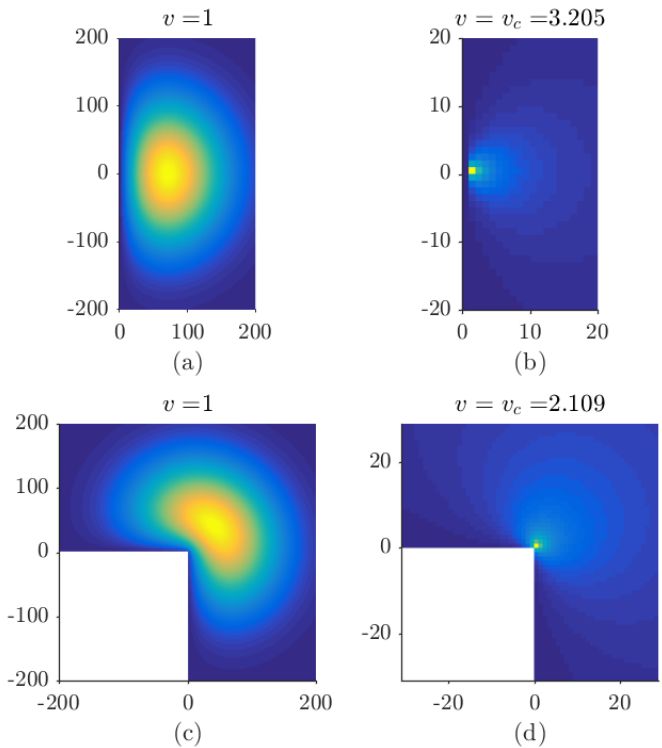}
\caption{
Probability density of a polymer end-point position ${\bf r}$
for $N=10^4$, near a flat surface [$\theta_0=\pi/2$, panels (a) and (b)],
and outside a rectangular wedge [$\theta_0=3\pi/4$, panels
(c) and (d)]. [See geometries of type ``1" in
Figs.~\ref{fig:DiscreteGeometries}(b) and (c).] In the absence of attraction
($v=1$) the distribution is broad with maximum at a distance $\sim\sqrt{N}$
away from the anchor point of the polymer [panels (a) and (c)], while at
$v=v_c$ [panels (b) and (d)] it is centered on the anchor point.
}
\label{fig:distribution}
\end{figure}

The presence of an adsorption transition can easily be  detected visually
by inspecting the probability of the end-point, proportional to $\tilde Z({\bf r},N)$.
In the absence of the attractive site ($v=1$), this probability density
is a Gaussian multiplied by a power law as in  Eq.~\eqref{eq:NdepSolution}.
Figures~\ref{fig:distribution}(a) and (c) depict such situations for geometries
of type ``1" in Figs.~\ref{fig:DiscreteGeometries}(b) and
\ref{fig:DiscreteGeometries}(c), respectively.
Even for a weakly attractive potential, i.e., for $v$ somewhat larger than 1,
the  distribution of the  polymer end-point approaches such a form for large $N$. At
the critical points $v=v_c$, shown in Figs.~\ref{fig:distribution}(b) and
\ref{fig:distribution}(d),
the distribution is still broad but remains centered on the attracting site to
which  the polymer is anchored, as expected from Eq.~\eqref{eq:NdepSolutionComp},
decaying as a power law cut-off at a distance of order $\sqrt{N}$. For
$v>v_c$ the distributions are still centered on the attracting site, but
the correlation length $\xi$ which cuts off the power-law decreases
with  increasing $v$, as will be discussed in the next Section.

\begin{figure}
\includegraphics[width=8 truecm]{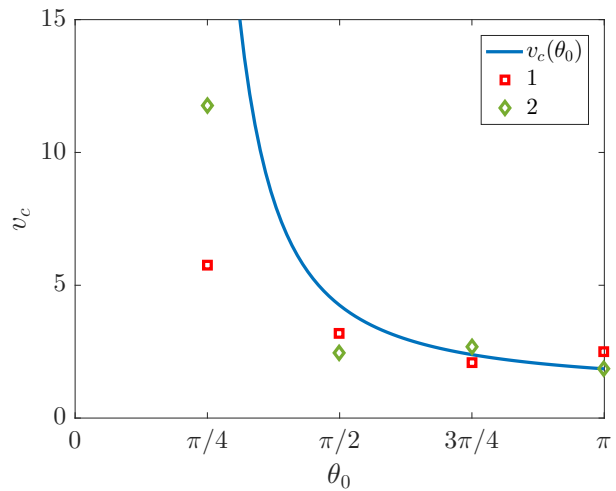}
\caption{Critical attractive strength $v_c$ for four values
of $\theta_0$, and for lattice realizations ``1" (squares) and ``2"
(diamonds) depicted in Fig.~\ref{fig:DiscreteGeometries}.
For comparison, the solid line shows the continuum result for a circular well,
with $v_c\equiv\exp(U_c/2d)$.}
\label{fig:v_c}
\end{figure}

We analyzed $\tilde{\cal Z}_{\rm tot}$ and $R^2$  for
all geometries depicted in
Fig.~\ref{fig:DiscreteGeometries}, obtaining a set of critical values
$v_c$ that  depend on both the opening semi-angle $\theta_0$, and on
the specific discrete realization of the wedge.
The values of $v_c$ for all eight cases are plotted
in Fig.~\ref{fig:v_c}. We expect $v_c$
to decrease with increasing $\theta_0$, and find results that qualitatively
resemble that of a continuous circular well. For each  value
of $\theta_0$ there is a difference between the two possible realizations
on a lattice, in variants denoted ``1" and ``2" in Fig.~\ref{fig:DiscreteGeometries},
with the variant that has more nearest-neighbor repulsive sites, $n$,
requiring a stronger attraction (larger $v_c$)  to confine the polymer.

\begin{figure}
\includegraphics[width=8.5 truecm]{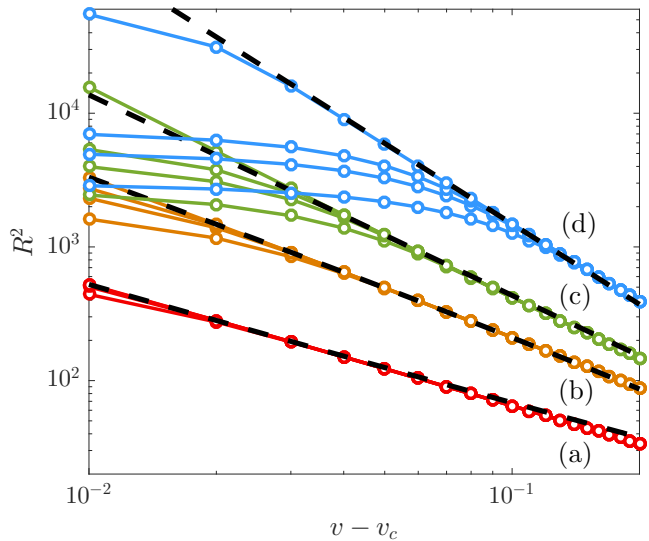}
\caption{
Logarithmic plots of the mean squared end-to-end distance $R^2$ as a function of
$v-v_c(\theta_0)$; different bundles of graphs correspond
to different $\theta_0$ realized by type ``1" lattice variants
in Fig.~\ref{fig:DiscreteGeometries} as indicated by the legends:
(a) $\pi/4$, (b) $\pi/2$, (c) $3\pi/4$.
and (d) $\pi$. Each bundle combines four different values of $N=4\cdot 10^3$, $7\cdot 10^3$,
$10^4$, and $10^5$ (bottom to top). The dashed lines represent the anticipated
power laws or power laws corrected by logarithms, as
explained in the text.}
\label{fig:Critical}
\end{figure}

\section{Critical behavior on a  lattice}\label{sec:LatticeCritical}

\begin{figure}[t!]
\includegraphics[width=8 truecm]{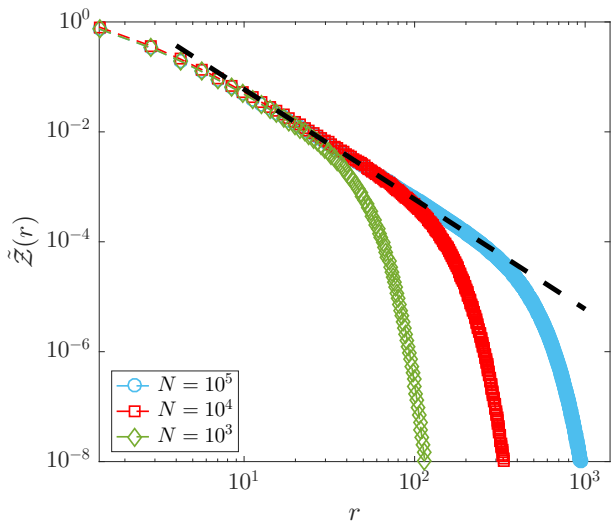}
\caption{
Logarithmic plot of the reduced partition function in type ``1" model
of Fig.~\ref{fig:DiscreteGeometries}(a) as a function of distance
$r$ along the diagonal, for (left to right) $N=10^3$, $10^4$ and $10^5$.
All curves have slope $-2$ for $r\ll\sqrt{N}$ as indicated
by the dashed line.
}
\label{fig:Zvsr}
\end{figure}

Figure~\ref{fig:Critical}
depicts on a logarithmic scale the numerically measured $R^2$ for several
values of $N$ as a function of $v-v_c$ close to the transition point. For
each $\theta_0$ we identify a linear segment corresponding to a power law
dependence of $R^2$, with possible logarithmic corrections. For large $N$
the power law regime is broader, and is cut off when $R^2$ reaches values
of order of $N$. The graphs represent rather different geometries of the
wedge, ranging from  a polymer confined inside a rectangular wedge
($\theta_0=\pi/4$), to outside a needle-like barrier ($\theta_0=\pi$).
Dashed lines indicate the theoretically predicted forms from the solution
of the continuous potential well in Eq.~\eqref{eq:xiVsT}. Note that, as in
Eq.~\eqref{eq:Rn}, we measure $R^2$ and not $\xi$. Thus for the polymer
inside a rectangular wedge with $m=2$, there is a logarithmic correction to
the relation between $R^2$ and $\xi^2$, leading to the expectation of $R^2\sim|(v-v_c)[c+\ln(v-v_c)]|^{-1}$. The fit in graph (a) in
Fig.~\ref{fig:Critical} uses $c=-4.6$. For the case of a flat line, as in
variant ``1" in Fig.~\ref{fig:DiscreteGeometries}(b), $m=1$ and the relation
between $R^2$ and $v-v_c$, has a logarithmic term  originating in the expression
for $\xi$ in Eq.~\eqref{eq:xiVsT}, resulting in  $R^2\sim |c+\ln(v-v_c)|/(v-v_c)$.
The fit in the graph (b) in Fig.~\ref{fig:Critical} uses $c=-1.6$. Finally,
the wedges of type ``1" in Figs.~\ref{fig:DiscreteGeometries}(c) and (d)
exhibit a simple power law scaling on $v-v_c$ with exponents of $-3/2$ and $-2$,
respectively, as shown in graphs (c) and (d) in Fig.~\ref{fig:Critical}.
The match between the numerical results and  theoretical predictions is quite
good, although the limited range of $N$ introduces some systematic errors:
for example, for the needle-like barrier,  direct measurement of the slope gives
$-1.95$ rather than $-2$, as the discreteness of the lattice combined with
limited $N$ introduces finite-size effects such as a slight effective reduction
of the angle $\theta_0$, and similar corrections of order $1/\sqrt{N}$.
We repeated the calculations also for variants of type ``2" in
Fig.~\ref{fig:DiscreteGeometries}, and despite the shifts in the positions of
$v_c$, the results were practically indistinguishable from those in
Fig.~\ref{fig:Critical}.

We further used the numerical results to check the validity of
Eq.~\eqref{eq:NdepSolutionComp} for the behavior of $\tilde{\cal Z}({\bf r_0},{\bf r})$
at the transition point $v_c$. Figure~\ref{fig:Zvsr} displays the weight of
polymers ending at a distance $r$ from the corner of a rectangular wedge. The
expected power-law scaling with exponent $-m=-2$ is clearly observed. [For proper
comparison one must replace $N'$ in the exponent in Eq.~\eqref{eq:NdepSolutionComp}
by $N/4$ where $N$ is the number of iterations used in the simulation.]
However, the curves for different $N$ do not show an increase as $N^{m-1}=N$,
for fixed $r\ll\sqrt{N}$, as predicted by Eq.~\eqref{eq:NdepSolutionComp}. Such
a power law increase is expected {\em exactly} at $v=v_c$, switching to a power-law
decrease as $N^{\gamma-1}=N^{-m/2}=N^{-1}$ for $v<v_c$. Thus, a small shift
in $v$ can change the behavior between power-law increase and power-law decrease.
In both cases there is no exponential growth with $N$ as occurs for $v>v_c$.
The simulations are not sufficiently sensitive to locate the exact position of
$v_c$. Indeed, the very choice of putative $v_c$ for Fig.~\ref{fig:Zvsr} was
guided by consideration that $\tilde{Z}_{\rm tot}$ is approximately constant over
the studied range of $N$,  likely leading to a value slightly below the true  $v_c$.

\section{Discussion}

In this work we studied the adsorption transition of a (phantom) polymer
to the corner of a (repulsive) wedge.
The scale free nature of the geometry leads to critical exponents
that depend on the opening semi-angle $\theta_0$ of the wedge:
At the transition point the probability density of the end point decays
as a power law $\sim r^{-m}$ with $m=\pi/2\theta_0$.
On approaching  the desorption point, the localization length  $\xi$
diverges with with an exponent of 1/2 for acute and obtuse angles,
then continuously increasing to one with increasing $\theta_0$ for reflex angles.
These results for $d=2$ are equally valid for an ideal polymer
near a wedge with an attractive edge in $d=3$. Once
self-avoidance is introduced, the two dimensional solution is no
longer applicable, since a self-avoiding polymer, as well as $\Theta$ polymer, cannot
be adsorbed to a finite volume. However, in $d=3$
the adsorption  transition of a self-avoiding and $\Theta$ polymer
to a wedge with adsorbing edge, as in Fig.~\ref{fig:PolymerOnWedge}, is expected to be
qualitatively similar, with albeit different $\theta_0$-dependent exponents.

The scale free geometry studied in this work combines objects of different dimensionality:
a zero-dimensional area of attraction with a one-dimensional repulsive surface.
In $d=3$ we can consider a richer class of scale-free objects (points, lines, planes, cones, pyramids, etc.),
and more combinations of zero-, one- and two-dimensional entities.
Each one of these components can be either repulsive or attractive, and we
expect to have competing and coexisting adsorption transitions. For $d=3$
the adsorption transition problem can be expanded to more realistic
self-avoiding and $\Theta$ polymers.

\begin{acknowledgments}
This work was supported by the National Science Foundation
under Grant No. DMR-1708280 (M.K.) and the Israel
Science Foundation Grant No. 453/17 (Y.K.).
\end{acknowledgments}

\end{document}